\documentclass{aa}

\begin{document}

\thesaurus{3 (11.19.2; 11.19.6; 11.11.1; 11.09.1 NGC 4062; 11.09.1 NGC 5248; 03.13.4)}

\title{The detection of spiral arm modulation in the stellar disk of \\
an optically flocculent and an optically grand design galaxy}

\author{Iv\^anio Puerari \inst{1},
        David L. Block  \inst{2},
        Bruce G. Elmegreen \inst{3},
        Jay A. Frogel \inst{4}
        \and
        Paul B. Eskridge \inst{4}}

\institute{Instituto Nacional de Astrof\'\i sica, Optica y Electr\'onica,
           Calle Luis Enrique Erro 1, 72840 Tonantzintla, Puebla, M\'exico
           \and
           Dept of Computational and Applied Mathematics, University Witwatersrand,
           Private Bag 3, WITS 2050, South Africa
           \and
           IBM Research Division, T.J. Watson Research Centre, P.O. Box 218,
           Yorktown Heights, NY 10598, USA
           \and
           Dept of Astronomy, The Ohio State University,
           140 W. 18th Avenue, Columbus, OH 43210-1173, USA}

\offprints{I. Puerari}

\date{Received .................. / Accepted ..................}

\titlerunning{Spiral arm modulation}
\authorrunning{I. Puerari et al.}

\maketitle

\begin{abstract}

Two dimensional Fourier spectra of near-infrared images of galaxies
provide a powerful diagnostic tool for the detection of spiral arm
modulation in stellar disks. Spiral arm modulation
may be understood in terms of interference patterns of outgoing and
incoming density wave packets or modes. The brightness along a spiral
arm will be increased where two wave crests meet and constructively
interfere, but will be decreased where a wave crest and a wave trough
destructively interfere. Spiral arm modulation has hitherto only been
detected in grand design spirals (such as \object{Messier 81}). Spiral
arm amplitude variations have the potential to become a powerful constraint
for the study of galactic dynamics.  We illustrate our
method in two galaxies: \object{NGC 4062} and \object{NGC 5248}.
In both cases, we have detected trailing and leading $m$=2 waves with similar
pitch angles. This suggests that the amplification mechanism is the WASER type
II. In this mechanism, the bulge region reflects (rather than refracts)
incoming waves with no change of pitch angle, but only a change of their sense
of winding.  The ratio between the amplitudes of the leading and the trailing
waves is about 0.5 in both cases, wherein the higher amplitude is consistently
assigned to the trailing (as opposed to leading) mode. The results are particularly
significant because \object{NGC 5248} is an optically grand design galaxy,
whereas \object{NGC 4062} is optically flocculent. \object{NGC 4062} represents
the very first detection of spiral arm modulation in the stellar disk of an
optically flocculent galaxy.


\end{abstract}

\section{Introduction}

Within the framework of the Density Wave Theory (Lin and Shu
\cite{linshu64}), the spiral arms of disk galaxies are the manifestation
of sets of travelling waves.  In the presence of both leading and
trailing sets of waves, the modal theory of galactic spiral structure
(Bertin et al. \cite{bertinetal89a}, \cite{bertinetal89b}) predicts that the
amplitude of a grand design two-arm spiral pattern will oscillate with radial
distance from the center because of the interference of wave packets or modes
which are propagating inward and outward, being reflected off a central bulge
(Lin \cite{lin70}; Lau et al. \cite{lauetal76}; Mark \cite{mark77};
Lin \cite{lin83}).

The swing amplification mechanism (Toomre \cite{toomre81}) may be
responsible for the appearance of the modes themselves.
The existence of trailing and leading waves is inferred by looking at the
unmistakable interference patterns on the density contours along spiral arms
(see, e.g., Toomre's Figures 10 and 12).  The presence of interference patterns
is betrayed in the discontinuity of the contours, or in other words, in the
modulation of the density (or surface brightness) along each arm. The
brightness along a spiral arm will be increased where two wave crests meet
and constructively interfere, but will be decreased where a wave crest and
a wave trough destructively interfere.

In the swing amplification models of Toomre (\cite{toomre81}),
the absence of a bulge or of a central concentration makes the
modelling unrealistic. The modes which grow in Toomre's models are
fast evolving.  In contrast, the modal theory of galactic spiral structure
assumes the presence of a central bulge. These bulges act as reflectors (often
termed the `Q-barrier'), resulting in a quasi-stationary modal pattern
(Thomasson et al. \cite{thomassonetal90}; Elmegreen and Thomasson
\cite{elmthom93}; Fuchs \cite{fuchs91, fuchs00}). Clearly, such models are
much more appropriate to the dynamics of spiral galaxies.

Several mechanisms have been proposed for the maintenance of spiral
structure in galaxies. The usual process of wave propagation, with
feedback and over-reflection, can maintain the wave pattern.
Mechanisms such as turbulent dissipation and shock formation in the
gaseous Population I component can also play a role in the self-regulation
of spiral modes (Bertin et al. \cite{bertinetal89a}, \cite{bertinetal89b}).

Symmetric spiral arm amplitude modulations indicative of underlying wave
modes have hitherto only been detected in the grand design galaxies
\object{M51}, \object{M81}, \object{M100} (Elmegreen et al.
\cite{elmelmsei89}) and in the multiple arm galaxy \object{M101}
(Elmegreen \cite{elmegreen95}). Arm variations in two other galaxies
were discussed by Grosb\o l (\cite{grosbol88}). M81 is one of the best studied
spiral galaxies, and the amplitude data is very useful to constrain the model
parameters within the modal theory.  The modal theory has been applied to
\object{M81} by Lowe et al. (\cite{loweetal94}), wherein the observed arm
modulation was modeled.

Near-infrared images reveal the old stellar Population II disk component
of spiral galaxies\footnote {Observational astronomers invariably
restrict the terminology of `Population II' for the
stars in the halo of a galaxy, and refer to  the `young Population I disk'
and the `old Population I disk'. When modelling the disks of galaxies,
however, theorists find it convenient to distinguish the two dynamically
different gaseous and stellar components by `gaseous Population I disk'
and `evolved stellar Population II disk', and we retain that terminology
here. It must, however, be emphasized that by an `old stellar Population II
disk' we are not referring to any true metal poor Population in the halo.},
while optical images show the rich variety of responses
of the young Population I component to the underlying older stellar
population responsible for the dynamics of the galaxies
(Frogel et al. \cite{frogeletal96}).
The young Population I disk component may only
constitute 5 percent of the dynamical mass of the disk of a galaxy.
For studying mass distributions of disk galaxies,
near-infrared images are essential (Block and Wainscoat \cite{blockwains91};
Block et al. \cite{blocketal94}; Quillen et al. \cite{aliceetal94},
\cite{aliceetal96}; Frogel et al. \cite{frogeletal96};
Block and Puerari \cite{davidivanio99}; Block et al. \cite{blocketal00}).

Optically thick dusty domains in galactic disks can completely
camouflage or disguise underlying stellar structures.
Dust extinction is highly effective whether or not the
dust lies in an actual screen or is well intermixed with the stars
(Elmegreen and Block \cite{elmdavid99}). The presence of dust and the
morphology of a galaxy are inextricably intertwined: indeed, the
morphology of a galaxy can completely change once the Population I
disks of galaxies are dust penetrated (e.g., Block
and Wainscoat \cite{blockwains91}; Block et al. \cite{blocketal94}).
Dust can completely obscure two armed grand design structure in some
optically flocculent galaxies (Thornley \cite{thornley96}; Block and
Puerari \cite{davidivanio99}; Elmegreen et al. \cite{elmetal99}).

In this paper, we propose a morphological method, based on the
bi-dimensional Fourier transform, to detect the existence of structures
with a different winding sense (trailing and leading patterns) in the same
galaxy. The galaxies for which spiral arm modulations  have hitherto
been detected (e.g., \object{M81}) are nearby.
The Fourier spectra offer an unambiguous way of identifying both
leading and trailing wave packets in galaxies which are not restricted
to be relatively close; the method can be applied to any spiral whose
stellar spiral arms are resolved.

The Fourier method is applied to the near-infrared images of two
galaxies which optically could not be more different: one is
flocculent (\object{NGC 4062}) whereas the other (\object{NGC 5248})
is grand design. The morphological appearances of \object{NGC 4062}
and \object{NGC 5248} in the dust penetrated regime are carefully
discussed below.

\section{Data and analysis}

The H band (1.65$\mu m$) image of \object{NGC 4062} is part of the OSU
(Ohio State University) Bright Spiral Galaxy Survey (Frogel et al.
\cite{frogeletal96}; Eskridge et al., in prep.). \object{NGC 5248}
was observed in the infrared (K$'$ 2.1$\mu m$) at the Observatorio
Astron\'omico Nacional at San Pedro Martir, Mexico, and forms part
of a larger project on deep K$'$ imaging. Details of reduction will
be given elsewhere.

Since our focus is morphology, no calibration frames are
required here. Using the IRAF\footnote{The IRAF
package is written and supported by the IRAF programming group at the
National Optical Astronomy Observatories (NOAO) in Tucson, Arizona.
NOAO is operated by the Association of Universities for Research in
Astronomy (AURA), Inc. under cooperative agreement with the National
Science Foundation (NSF).} task IMEDIT,
the images were cleaned of any foreground stars. The galaxies were
then deprojected using the IRAF ROTATE and MAGNIFY routines.
Deprojection parameters (position angle PA and inclination $w$)
are listed in Table \ref{tabparm}; also given in that Table
are van den Bergh luminosity classes and blue absolute magnitudes
as determined by Sandage and Tammann (\cite{sandtamm87}).
Morphological parameters such as DP and arm classes
are discussed below.

\begin{table}
\caption[]{Parameters of the galaxies}
\label{tabparm}
\begin{tabular}{lcc}

  & \object{NGC 4062} & \object{NGC 5248} \\

\hline

Type$^a$ & Sc(s)II-III & Sbc(s)I-II \\

M$^a_{B_T}$ & $-$19.44 & $-$21.19 \\

DP class$^b$ & E$\beta$ & E$\beta$ \\

arm class$^c$ & 3 & 12 \\

PA$^d$ & 100$^{\circ}$ & 110$^{\circ}$ \\

$w^d$ & 64$^{\circ}$ & 43$^{\circ}$ \\

Filter & H (1.65$\mu m$) & K$'$ (2.1$\mu m$) \\

\hline

\end{tabular}

$^a$ Sandage and Tammann \cite{sandtamm87}

$^b$ Block and Puerari \cite{davidivanio99}

$^c$ Elmegreen and Elmegreen \cite{elmelm87}

$^d$ de Vaucouleurs et al. \cite{rc3}

\end{table}

Once the galaxies are corrected to a `face-on'
orientation, we applied the program 2dfft (see the appendix of
Schr\"oeder et al.  \cite{schroederetal94}), which calculates the fast
Fourier transform of a given image using a basis of logarithmic spirals.
As shown in an extensive work by Danver (\cite{danver42}), logarithmic
spirals appear to be the best {\it mathematical} description for
galactic arms. In a more recent work, Kennicutt (\cite{kenni81})
concluded that logarithmic, as well as hyperbolic
spirals, are good representations of galactic spiral arms.
As discussed elsewhere (e.g., Consid\`ere and Athanassoula
\cite{consathan82}; Puerari and Dottori \cite{puedot92}),
the choice of logarithmic spirals does
not constrain the analysis. They only form the basis in a vector space for the
decomposition.  If only a few coefficients -- typically, one or two -- are
required to reconstruct the original image using inverse Fourier
transforms (as is the case here), the choice of logarithmic spirals is
indeed appropriate.

The Fourier method has been extensively discussed in a number of papers (e.g.,
Kalnajs \cite{kalnajs75}; Consid\`ere and Athanassoula \cite{consathan82}; Iye
et al. \cite{iyeetal82}; Puerari and Dottori \cite{puedot92};
Puerari \cite{puerari93}, amongst others). In the Fourier method, an image is
decomposed into a basis of logarithmic spirals
of the form $r$=$r_o {\rm exp} (-{m\over p} \theta)$.
The Fourier coefficients $A(p,m)$ can be written as

$$ A(p,m) = \frac{1}{D}\int_{-\pi}^{+\pi}\int_{-\infty}^{+\infty} I(u,\theta)
{\rm exp}[-i(m \theta + p u)] d u d \theta$$

\begin{figure*}
     \vspace{9.55cm}
          \includegraphics{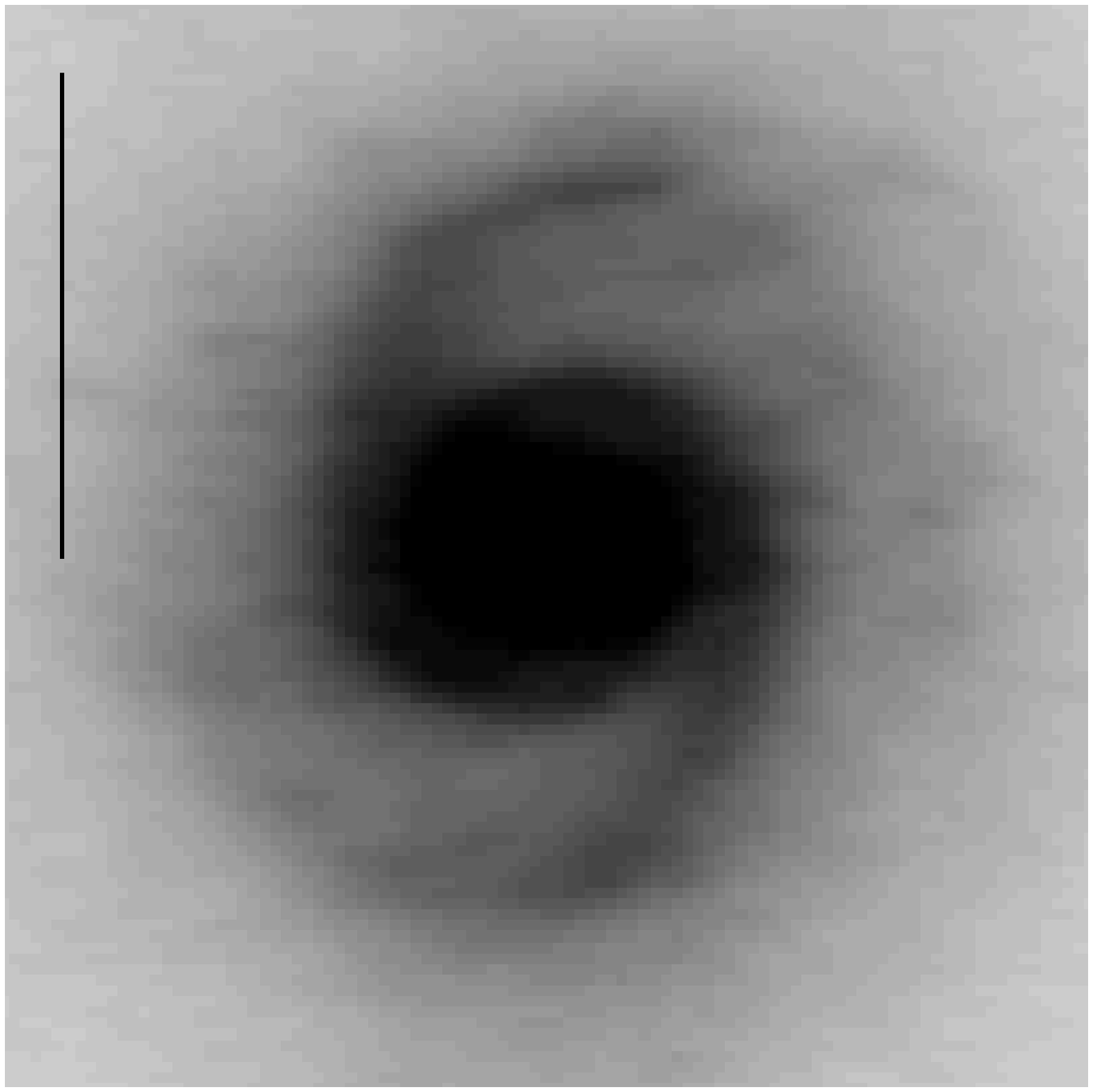}
          \includegraphics{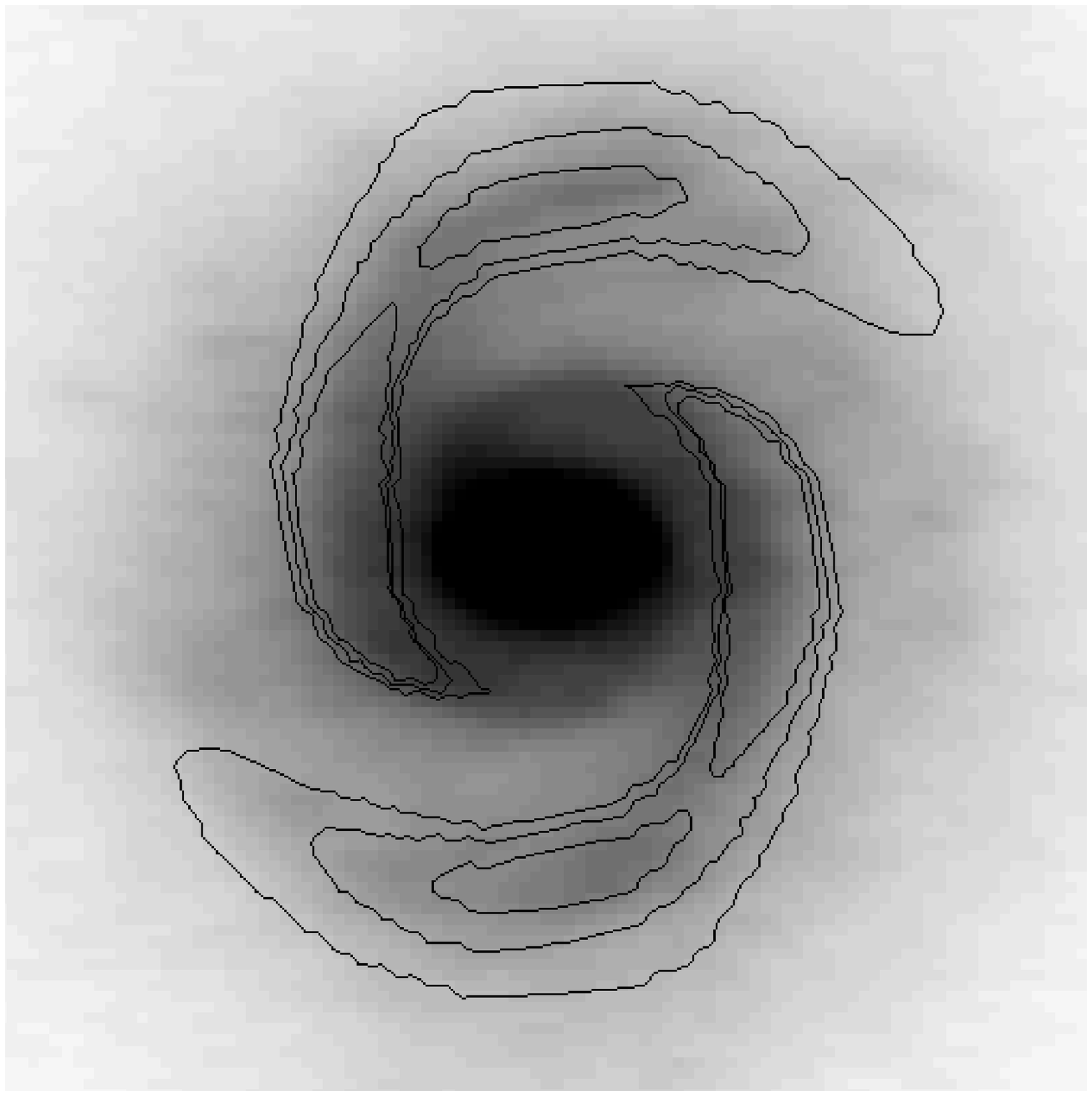}
          \caption{Left: Deprojected H image of \object{NGC 4062}. The vertical
             bar represents 5 kpc (H$_0$=50 km/sec/Mpc). Right: The contours of
             the inverse Fourier transform for the $m$=2 mode are overlayed on
             the deprojected image. The contours are the real part of the complex
             spatial function $S_2(u,\theta)$=$S_2(u)e^{i2\theta}$. The lowest
             and the highest plotted contours represent 3\% and 30\% of the
             maximum amplitude for the $m$=2 mode, respectively.}
\label{imag4062}
\end{figure*}

Here, $u \equiv {\rm ln}\;r$, $r$ and $\theta$ are the polar coordinates,
$m$ represents the number of the arms, $p$ is related to the pitch angle
$P$ of the spiral by  $P$=${\rm atan}(-m/p)$, and $I(u,\theta)$ is the
distribution of light of a given deprojected galaxy, in a
\hbox{(${\rm ln}\;r$, $\theta$)} plane. $D$ is a normalization factor
written as

$$ D=\int_{-\pi}^{+\pi} \int_{-\infty}^{+\infty} I (u, \theta) d u d \theta $$

In practice, the integrals in $u \equiv {\rm ln}\;r$ are
calculated from a minimum radius (selected to exclude the bulge where there
is no information of the arms) to a maximum radius (which extends to the outer
limits of the arms in our images).

The inverse Fourier transform can be written as

$$S(u,\theta) = \sum_m S_m (u) {\rm e}^{im \theta} $$

\noindent where

$$ S_m(u) = \frac{D}{{\rm e}^{2u} 4 \pi^2} \int_{p_-}^
{p_+}\;G_m (p) A(p,m) {\rm e}^{i p u}\;dp$$

\noindent and $G_m(p)$ is a high frequency filter used to smooth
the $A(p,m)$ spectra at the interval ends (see Puerari and Dottori
\cite{puedot92}), and it has the form

$$ G_m(p) = {\rm exp} \left[ -\frac{1}{2} \left( \frac{p - p_{max}^m}{25}
\right)^2 \right] $$

\noindent where $p_{max}^m$ is the value of $p$ for which the amplitude of
the Fourier coefficients for a given $m$ is maximum. The chosen interval ends
($p_+$=$+50$ and $p_-$=$-50$), as well as the step-size $dp$=$0.25$, are
suitable for the analysis of galactic spiral arms.

In Table \ref{tabspec}, we present the values for the dominant $m$=2
components in the Fourier spectra. In that Table, $p_{max}$ is the
value where the spectrum for $m$=2 peaks, $\vert A(p_{max},2)\vert$
is the amplitude of the peak, and $\vert A^L \vert/\vert A^T \vert$
denotes the corresponding ratio between the leading ($L$) and
the trailing ($T$) amplitudes.  $P$ is the pitch angle of the spiral,
related to $p_{max}$ by  $P$=${\rm atan}(-2/p_{max})$. $\Phi$ is the
phase of the spiral arm, calculated as $\Phi$=${\rm atan}(Im [A]/Re [A])$,
where $Im [A]$ and $Re [A]$ are the imaginary and the real part of
$A(p_{max},2)$, respectively. A diagram explicitly showing the
definition of trailing as opposed to leading spiral arms may be
seen in Figure 15 of Athanassoula (\cite{athan84}), in Figure
2.8 of Bertin and Lin (\cite{bertinlin96}) or in Figure 6.5 of
Binney and Tremaine (\cite{binneytremaine87}).

\begin{table}
\caption[]{Values of the $m$=2 Fourier spectrum}
\label{tabspec}
\begin{tabular}{lcc}

  & \object{NGC 4062} &  \object{NGC 5248} \\

\hline

$p_{max}^T$ & $-$5.75 & $-$3.75 \\

$p_{max}^L$ & 4.5 & 3.5 \\

$\vert A(p_{max}^T,2)\vert$ & 2.1E$-$3 & 1.5E$-$1 \\

$\vert A(p_{max}^L,2)\vert$ & 1.2E$-$3 & 8.0E$-$2 \\

$\vert A^L \vert/\vert A^T \vert$ & 0.54 & 0.54 \\

$P^T$ & 19$^\circ$ & 28$^\circ$ \\

$P^L$ & $-$24$^\circ$ & $-$30$^\circ$ \\

$\Phi^T$ & $-$34$^\circ$ & $-$87$^\circ$ \\

$\Phi^L$ & $-$75$^\circ$ & $-$72$^\circ$ \\

\hline

\end{tabular}

\end{table}

It is evident from Table \ref{tabspec} that the pitch angle for the
trailing and the leading components for each galaxy is very similar.
A classification scheme of spiral galaxies in the near-infrared
was recently proposed by Block and Puerari (\cite{davidivanio99}).
Galaxies are binned into three groups $\alpha$, $\beta$ and
$\gamma$ based on the pitch angle of the arms, robustly determined
from Fourier spectra.  Even-sided (as opposed
to lopsided) galaxies have a dominant $m$=2 component in the
Fourier spectra, and they are designated in this scheme by an
`E'. \object{NGC 4062} and \object{NGC 5248} both belong
to the dust penetrated E$\beta$ class (Table \ref{tabparm}).

\begin{figure}
     \vspace{9.0cm}
          \includegraphics{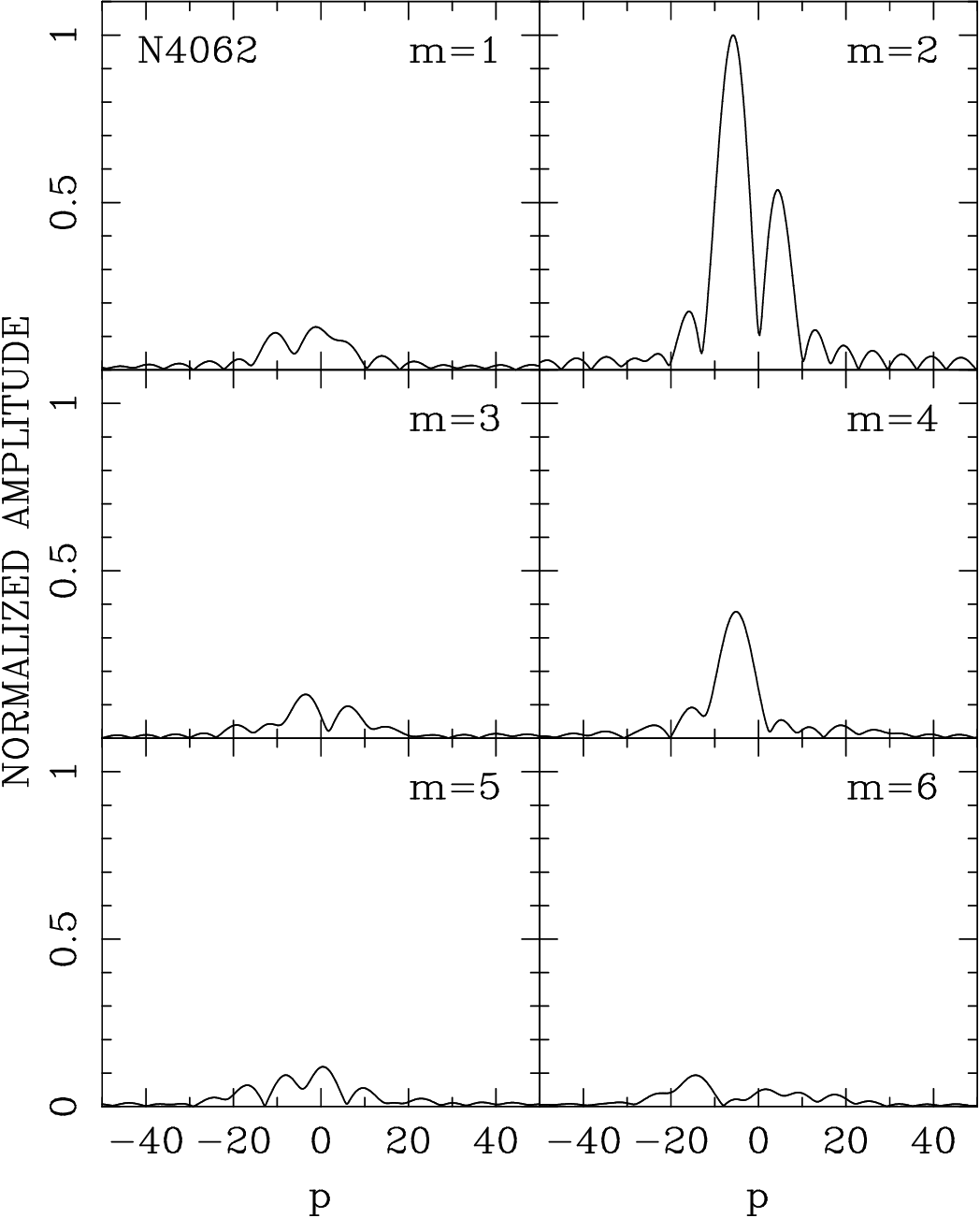}
          \caption{Fourier spectra for \object{NGC 4062}. Note the absence of odd
             modes which is indicative of a highly bisymmetrical light distribution.}
\label{spec4062}
\end{figure}

The ratio between the
amplitudes of the leading and the trailing patterns is also the same
for the two galaxies. This is particularly interesting, since there
is a large difference in the linear size of the two galaxies
(adopting H$_0$=50 km/sec/Mpc, the linear diameters of \object{NGC 4062}
and \object{NGC 5248} are 18 and 42 kpc, respectively).
Note furthermore that the ratio of amplitudes does not depend
on the absolute magnitude of the parent spiral,
neither on the arm class (see Table \ref{tabparm}).

In the determination of Fourier coefficients and of pitch angle,
careful deprojections to face-on are of course necessary.
Mean uncertainties of position angle and inclination angle as
a function of inclination are drawn in
Fig. 2 of Consid\`ere and Athanassoula (\cite{consathan88}).
For \object{NGC 4062}, we find (using eight
deprojected runs, wherein inclination and position angle are
systematically varied) that incorrect
deprojections can introduce a maximum difference of 13\% in the reported
$\vert A^L\vert$/$\vert A^T\vert$ ratio (see Table \ref{tabspec}). For
the pitch angle, the maximum difference we find is only 5$^\circ$.
The situation is slightly more complex for \object{NGC 5248},
because of
its smaller inclination (and hence, larger uncertainties in the
deprojection
angles) and its asymmetry. For a few incorrect deprojections (mainly
when we deproject the image with both an incorrect PA and an
incorrect $\omega$),
the leading component does not appear clearly. In other cases, the
errors in the $\vert A^L\vert$/$\vert A^T\vert$ ratio and in the
pitch angles are of the order of the errors listed for \object{NGC 4062}.
The results are fully consistent with the findings of Block et al.
(\cite{blocketal00}) wherein determination of pitch
angles from Fourier spectra are found to be surprisingly robust;
galaxies do not move from one dust penetrated (DP) class to the next.
\object{NGC 4062} and \object{NGC 5248} remain dust penetrated $\beta$ class.

\subsection{\object{NGC 4062}}

Optical images of the galaxy \object{NGC 4062} (Sandage and Bedke
\cite{sandagebedke94}, plate 265) reveal numerous patches of star
formation with no grand design spiral structure from a density wave
in the underlying stellar disk. Such patchy structure occurs in over
sixty percent of isolated, non-barred galaxies, giving them a flocculent,
fleece-like appearance (Elmegreen and Elmegreen \cite{elmelm87}). A
characteristic of flocculent galaxies such as \object{NGC 4062} is that the
optical patches, by definition, span only a small range in azimuth.

\begin{figure*}
     \vspace{9.5cm}
          \includegraphics{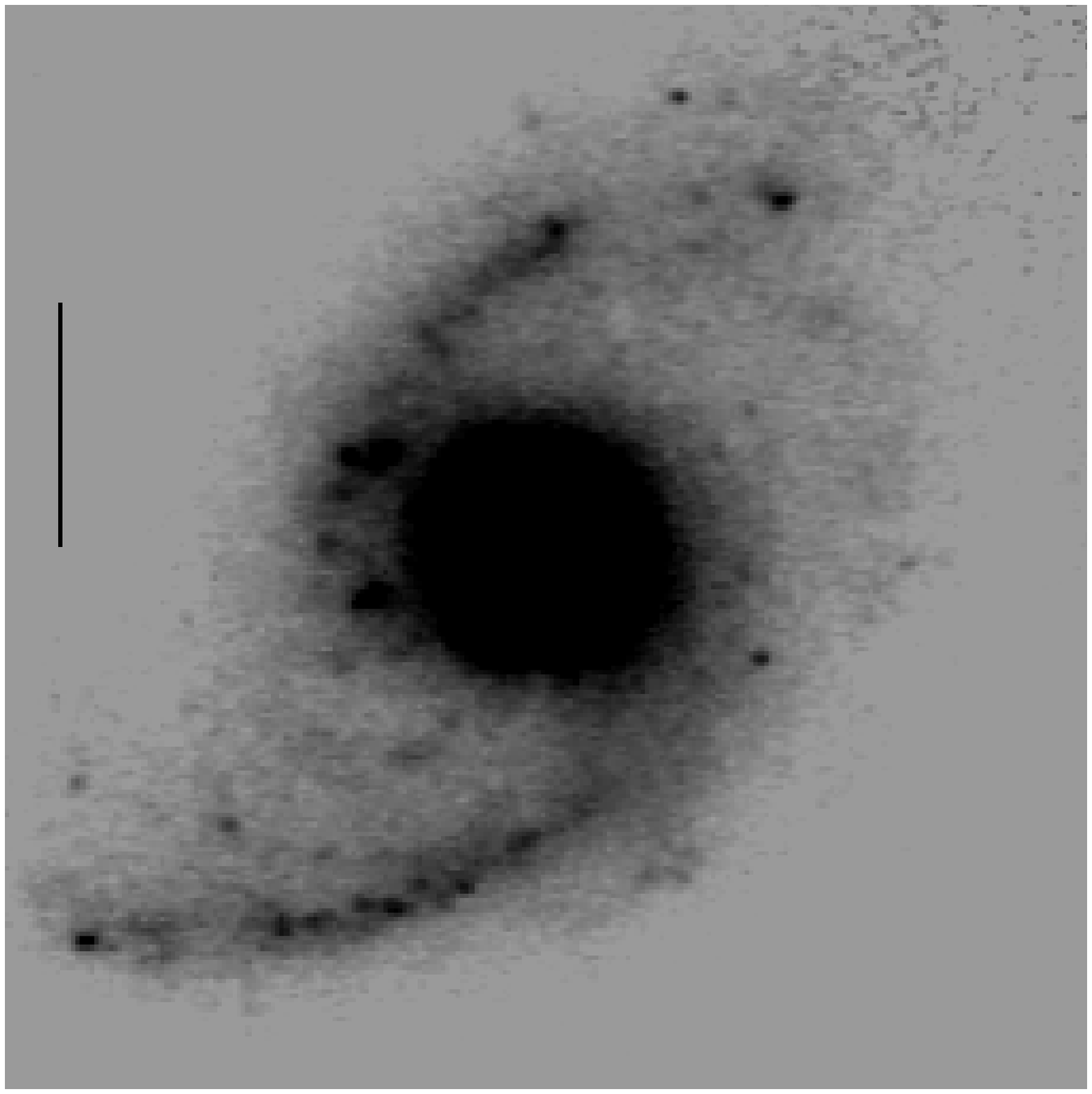}
          \includegraphics{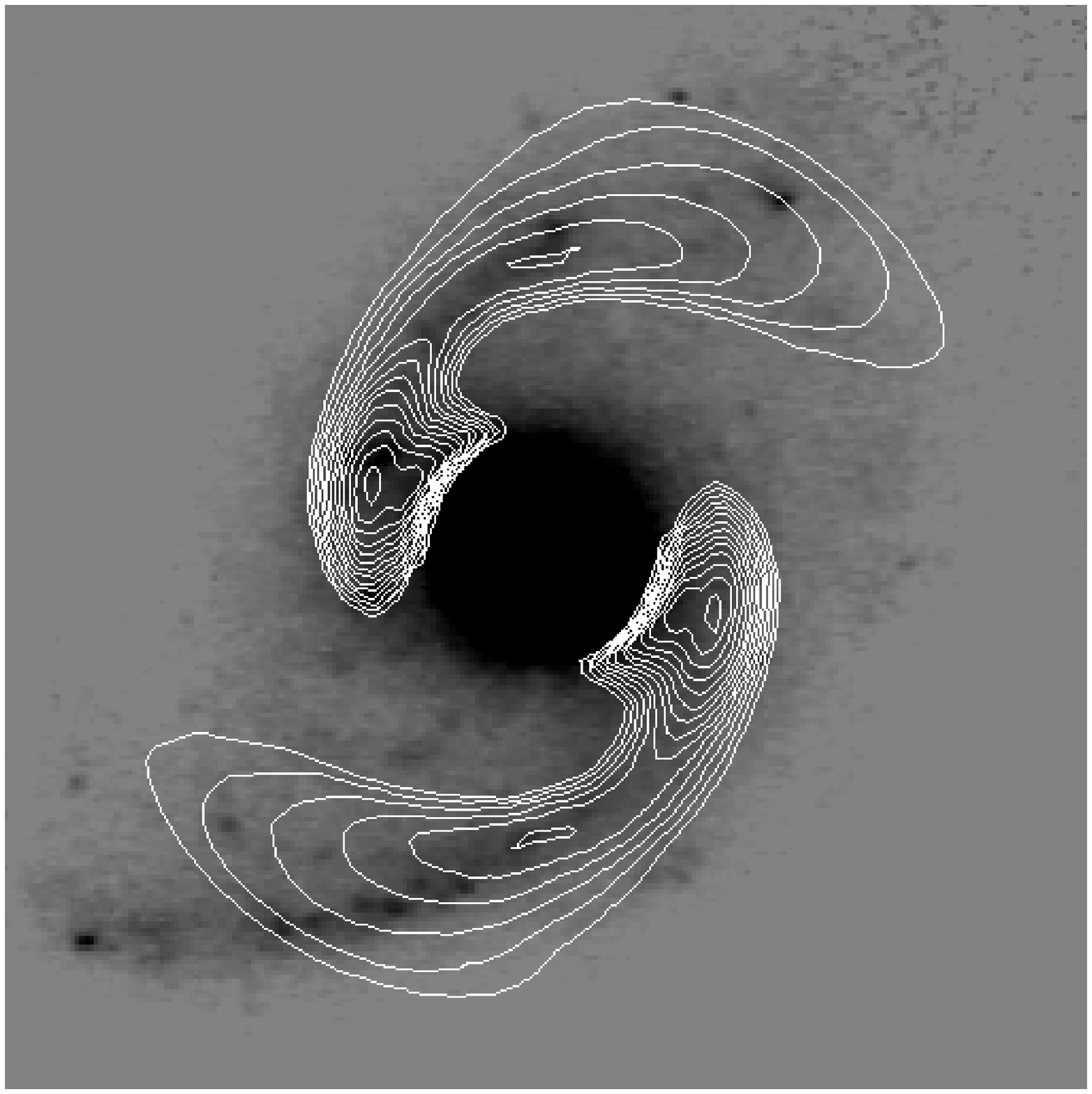}
          \caption{Left: Deprojected K$'$ image of \object{NGC 5248}.
             The vertical bar represents 5 kpc (H$_0$=50 km/sec/Mpc).
             Right: The contours of the inverse Fourier transform for
             the $m$=2 mode are overlayed on the deprojected image. The
             lowest and the highest plotted contours represent 8\% and
             90\% of the maximum amplitude for the bisymmetric mode,
             respectively.}
\label{imag5248}
\end{figure*}

\object{NGC 4062} belongs to arm class 3, described as `fragmented arms
uniformly distributed around the galactic center' (see  Elmegreen and
Elmegreen \cite{elmelm87}). What is so striking in the dust penetrated,
near-infrared regime is that \object{NGC 4062} presents a remarkable, bisymmetrical
grand design morphology (see Figure \ref{imag4062}): two arms spanning over 90
degrees in azimuth. It is evident that the young Population I and old stellar
Population II disks of \object{NGC 4062} actually decouple\footnote{What is meant by
decoupling is that a galaxy may show two different morphologies
when examined optically and in the near-infrared regime. For example,
\object{NGC 309} is classified as Sc in the optical, but appears as a SBa
in the near-infrared (Block and Wainscoat \cite{blockwains91}).
Two different morphologies in the same galaxy may co-exist via a
feedback mechanism or dynamical thermostat (Bertin and Lin
\cite{bertinlin96}).}.

Another surprise is that the Fourier spectra of \object{NGC 4062} do not present a
single peak for the $m$=2 mode (Figure \ref{spec4062}). The presence of
two peaks betrays the existence of two different spirals. One would have
the form of an `Z' (coming from the peak with maximum at $p<0$) and the
other having a `S' form. This situation is exactly what is expected
within the framework of the swing amplification theory or in
the modal theory of galactic spiral structure: a strong
trailing pattern and a weaker leading one.

\subsection{\object{NGC 5248}}

Grand design spirals (where two dominant arms may span many
degrees in azimuth) are completely different from flocculent galaxies.
Possibly two of the best studied grand design spirals are \object{Messier 81}
(\object{NGC 3031}) and \object{M51} (\object{NGC 5194}), where
two long symmetric arms dominate their optical disks.
Elmegreen and Elmegreen (\cite{elmelm87}) devote classes 9 through 12 to
the grand design bin.

\object{NGC 5248} is a magnificent, grand design optical specimen. Its arm
class is 12 (the same class to which \object{M81} and \object{M51} belong). In grand
design galaxies, density waves are believed to have organized young
stellar associations to form the symmetric optical spiral pattern
(Elmegreen \cite{elmegreen95}) and one often finds a strong coupling
between the young Population I and the older stellar Population II disks.
This is the case for \object{NGC 5248}.

\begin{figure}
     \vspace{9.0cm}
          \includegraphics{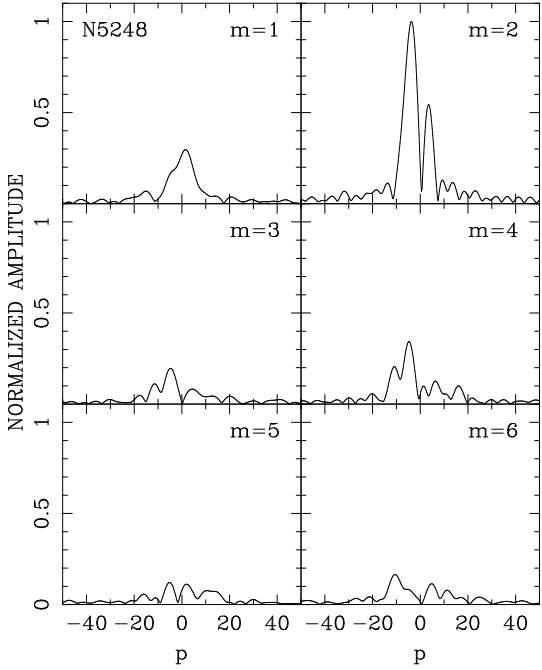}
          \caption{Fourier spectra for \object{NGC 5248}. The peak at $m$=1
              represents the asymmetry of the galaxy (see Figure \ref{imag5248}).
              Note the striking similarity between the spectrum for $m$=2 in this
              Figure and that in Figure \ref{spec4062}.}
\label{spec5248}
\end{figure}

This galaxy, however, is not quite as symmetric as \object{NGC 4062} is in the
near-infrared K$'$ image (see Figure \ref{imag5248}). The two trailing arms
have a small difference in their winding angle, with one arm being slightly
more open than the other one. In the Fourier spectra, $m$=1
components reveal this asymmetry, and this is attested to by the
relatively large $m$=1 component of \object{NGC 5248} (Figure \ref{spec5248}).

Nevertheless, the Fourier spectra in Figure \ref{spec5248}
reveal something quite remarkable: the old stellar Population II
disk of an optically grand design galaxy, \object{NGC 5248}, can be almost
identical to that of an optically flocculent, \object{NGC 4062} (compare Figures
\ref{spec5248} and \ref{spec4062}).  \object{NGC 5248} also has two peaks for
the $m$=2 mode, one at $p<0$ and another for $p>0$,
which proves the existence of two wave-trains with different
winding senses: one trailing, the other leading.

\subsection{The trailing and the leading spirals}

In order to shed further information about the modes we have detected,
we calculated the inverse Fourier transform in a different way to that
illustrated in Figs. \ref{imag4062} and \ref{imag5248}. The goal is to
separate the trailing and the leading components, and in order to
study 
their individual 
characteristics,  we calculate the $S_2(u)$
functions by considering only $p<0$ or $p>0$. In other words, 
for the trailing mode,
we use

$$ S^T_m(u) = \frac{D}{{\rm e}^{2u} 4 \pi^2} \int_{p_-}^
{0}\;G_m (p) A(p,m) {\rm e}^{i p u}\;dp$$

\noindent and for the leading one,

$$ S^L_m(u) = \frac{D}{{\rm e}^{2u} 4 \pi^2} \int_{0}^
{p_+}\;G_m (p) A(p,m) {\rm e}^{i p u}\;dp$$

\begin{figure}
     \vspace{6.75cm}
          \includegraphics{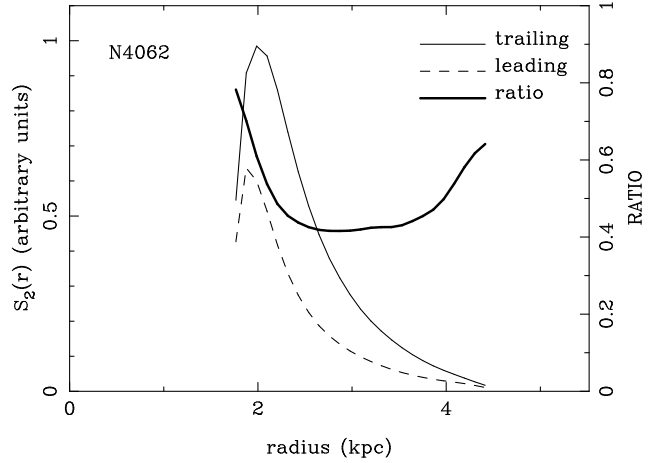}
          \caption{The surface density $S^L_2(r)$ and $S^T_2(r)$ functions of
             \object{NGC 4062}, and the ratio $S^L_2/S^T_2$. The cuts in radius
             are the minimum and maximum radii used in the Fourier analysis.}
\label{surface4062}
\end{figure}

\begin{figure}
     \vspace{6.75cm}
          \includegraphics{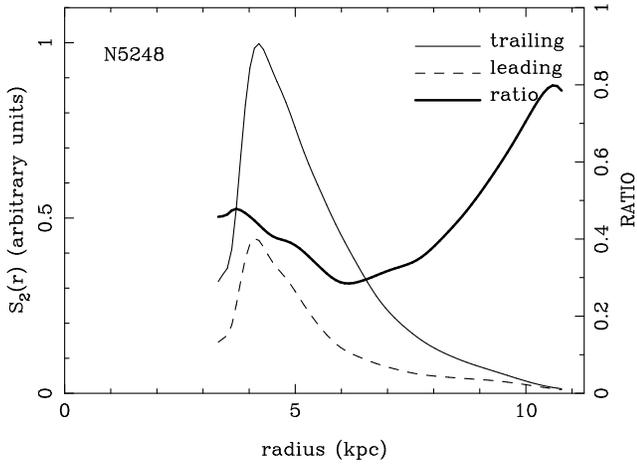}
          \caption{Same as Fig. \ref{surface4062}, but for \object{NGC 5248}.}
\label{surface5248}
\end{figure}

In Figs. \ref{surface4062} and \ref{surface5248} we plot the $S^L_2(u)$
and the $S^T_2(u)$ functions for \object{NGC 4062} and \object{NGC 5248},
respectively. These functions peak where we find the first maximum in
the arm/interarm contrast (see Figs. \ref{armintarm4062} and \ref{armintarm5248}
below). Also plotted in Figs. \ref{surface4062} and
\ref{surface5248}
is the ratio between the $S^L_2(u)$ and the $S^T_2(u)$ functions. This
ratio
is important for comparisons with theoretical studies. As one can see,
the
ratio is of the order of 0.4$-$0.5 for almost all the spatial extent of
the spiral
arms in both galaxies. The ratio shows an increase towards smaller
distances (nearer to the bulges of the galaxies) and also for larger
radii (probably near to the co-rotation radius - see below).

\begin{figure}
     \vspace{6.5cm}
          \includegraphics{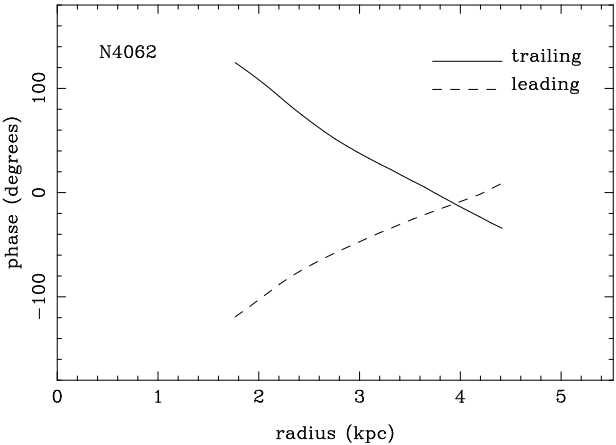}
          \caption{The phase of the leading and the trailing modes of
             \object{NGC 4062}.}
\label{phase4062}
\end{figure}

\begin{figure}
     \vspace{6.5cm}
          \includegraphics{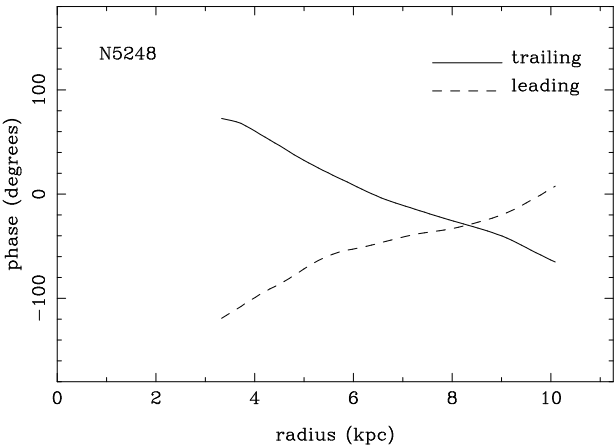}
          \caption{Same as Fig. \ref{phase4062}, but for \object{NGC 5248}.}
\label{phase5248}
\end{figure}

By using the $S^L_2(u)$ and the $S^T_2(u)$ functions, we can
separate the leading and the trailing spirals of each galaxy.
In Figs. \ref{phase4062} and \ref{phase5248} we plot the phase of each component
(see Puerari and Dottori \cite{puedot97}) for \object{NGC 4062} and
\object{NGC 5248}, respectively. The phase represents the azimuthal
position of the maximum intensity of each spiral. So, when the difference between
the phases of the leading and the trailing components is 0$^{\circ}$ or
180$^{\circ}$, the spiral pattern of the galaxy must show a maximum. In
contrast, when the phase difference is 90$^{\circ}$, the spiral arms
of the galaxy must show a minimum. For \object{NGC 4062}, we find
2.2 and 3.9 kpc for the positions of the maxima, and 2.9 kpc for the
minimum. These values are in complete agreement with the peaks and
the dip in the arm/interarm contrast (see Fig. \ref{armintarm4062},
below).
The situation is a little more complex for \object{NGC 5248}. For
this
galaxy, the values for the positions of the maxima are 3.6 and 8.3 kpc, and
for the minimum, we find 5.2 kpc. These values do not fit the
arm/interarm 
contrast as well as does the data for \object{NGC 4062} (see 
Fig. \ref{armintarm5248}, below). This discrepancy could be explained
by the fact that \object{NGC 5248} is not quite as symmetric as \object{NGC 4062}.

Future models might be able to use these leading and trailing waves
to locate the co-rotation radius. Leading modes propagate outward inside
co-rotation and inward outside co-rotation, so that a leading spiral
commencing from the nuclear region would only extend to the co-rotation
radius; in other words, co-rotation would be where the leading spiral
ends (C. Yuan, private communication). The difficulty in the
present study in placing the co-rotation resonance is that the leading arm
is not directly seen in our near-infrared images; rather, its existence is
inferred from the arm modulation.


\subsection{Arm/Interarm contrast}

It is well known that the arm/interarm contrast in spiral
galaxies usually increases with radius (Elmegreen and
Elmegreen \cite{elmelm84}). While some of the galaxies
in their sample revealed arm/interarm profiles with a simple
sinusoidal pattern in the I band (0.85$\mu m$), a large percentage of
them showed a chaotic behaviour. Some galaxies can be
optically thick even at I. We have calculated the arm/interarm contrast
following Elmegreen and Elmegreen (\cite{elmelm84}). We have drawn azimuthal
profiles for a number of radii, and by using an interactive plotting program,
we have taken a mean value at the maxima (where the spiral arms are located)
and a mean value at the `troughs' (the interarm regions). The final value is
calculated from intensity values as follows:

$${\rm arm/interarm=(arm-interarm)/0.5(arm+interarm)}$$

\begin{figure}
     \vspace{6.5cm}
          \includegraphics{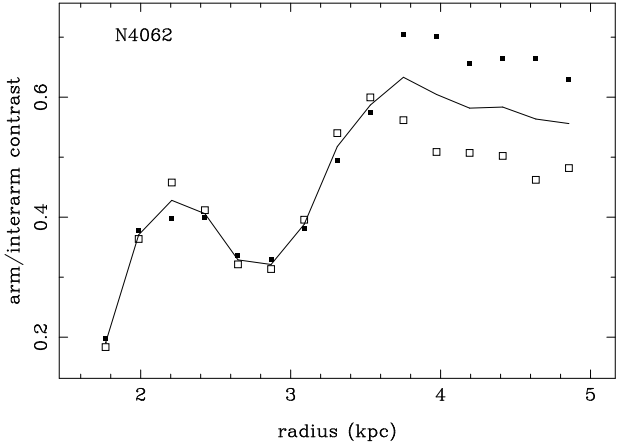}
          \caption{Arm/interarm contrast for the H band image of
             \object{NGC 4062}. Open and filled squares are the
             measures in the two arms. The solid line is the mean
             of the two values at each radius.}
\label{armintarm4062}
\end{figure}

\begin{figure}
     \vspace{6.5cm}
          \includegraphics{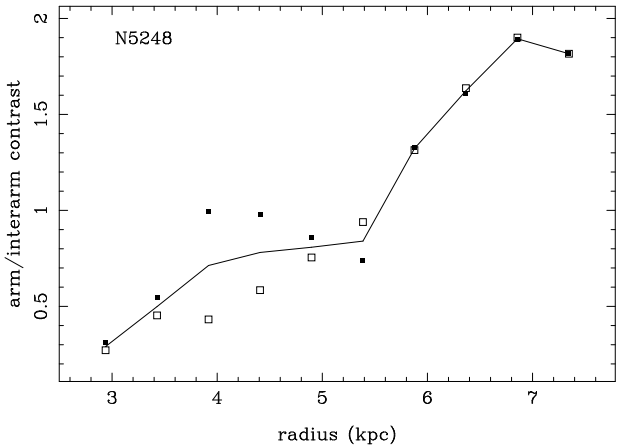}
          \caption{Same as Figure \ref{armintarm4062}, but for
             the K$'$ image of \object{NGC 5248}.}
\label{armintarm5248}
\end{figure}

In Figures \ref{armintarm4062} and \ref{armintarm5248} we plot the
calculated arm/interarm contrast for \object{NGC 4062} and \object{NGC 5248},
respectively. The maximum values for \object{NGC 4062} (the flocculent
galaxy) are approximately 3 times less than those for \object{NGC 5248}
(grand design). As one can see, the arm/interarm contrast increases
with radius, but not in a monotonic way. The modulation of the
intensity is caused by the interference between the incoming and the
outgoing density waves and is clearly evident in our plots. The sinusoidal
behaviour is even more remarkable in the flocculent \object{NGC 4062}.

It is important to note that the values for the
arm/interarm contrast calculated
for these two galaxies are in complete agreement with other studies (eg.,
Elmegreen et al. \cite{elmegreenetal96}). Values of arm/interarm contrast
as high as 2 are not unreasonable (see Fig. 6 of Elmegreen et al.
\cite{elmegreenetal96}). A near-infrared study of M100 by Gnedin et
al. (\cite{gnedinetal96}) yields an arm/interarm contrast of 3. 
High arm/interarm contrasts do not imply
unrealistic high perturbations of the velocity fields. Rather, what
is important
is the ratio of the arm mass to the enclosed galaxy mass at a given
radius. The arm is but a small fraction of the entire mass of the galaxy
inside
any given radius, so that the streaming motions are small, even for
high arm/interarm contrasts.


\begin{figure}
     \vspace{9.5cm}
          \includegraphics{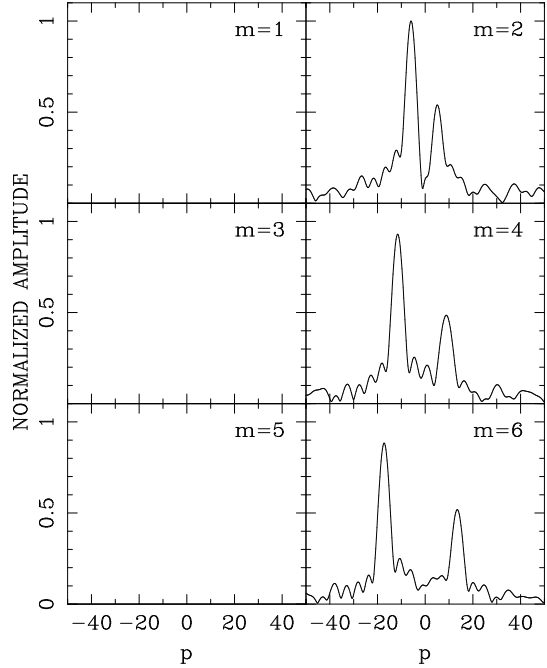}
          \caption{Fourier spectra for synthetic spirals constructed with
             the parameters of \object{NGC 4062} (Table 2). The strong harmonics
             at $m$=4 and $m$=6 came because we have no dispersion of the
             `density' around the mathematical logarithmic spirals
             (see discussion in Puerari and Dottori 1992). The synthetic
             spirals are {\sl exactly} bisymmetrical, and thus the amplitude of
             the odd modes is zero.}
\label{specmodel}
\end{figure}

\begin{figure}
     \vspace{9.5cm}
          \includegraphics{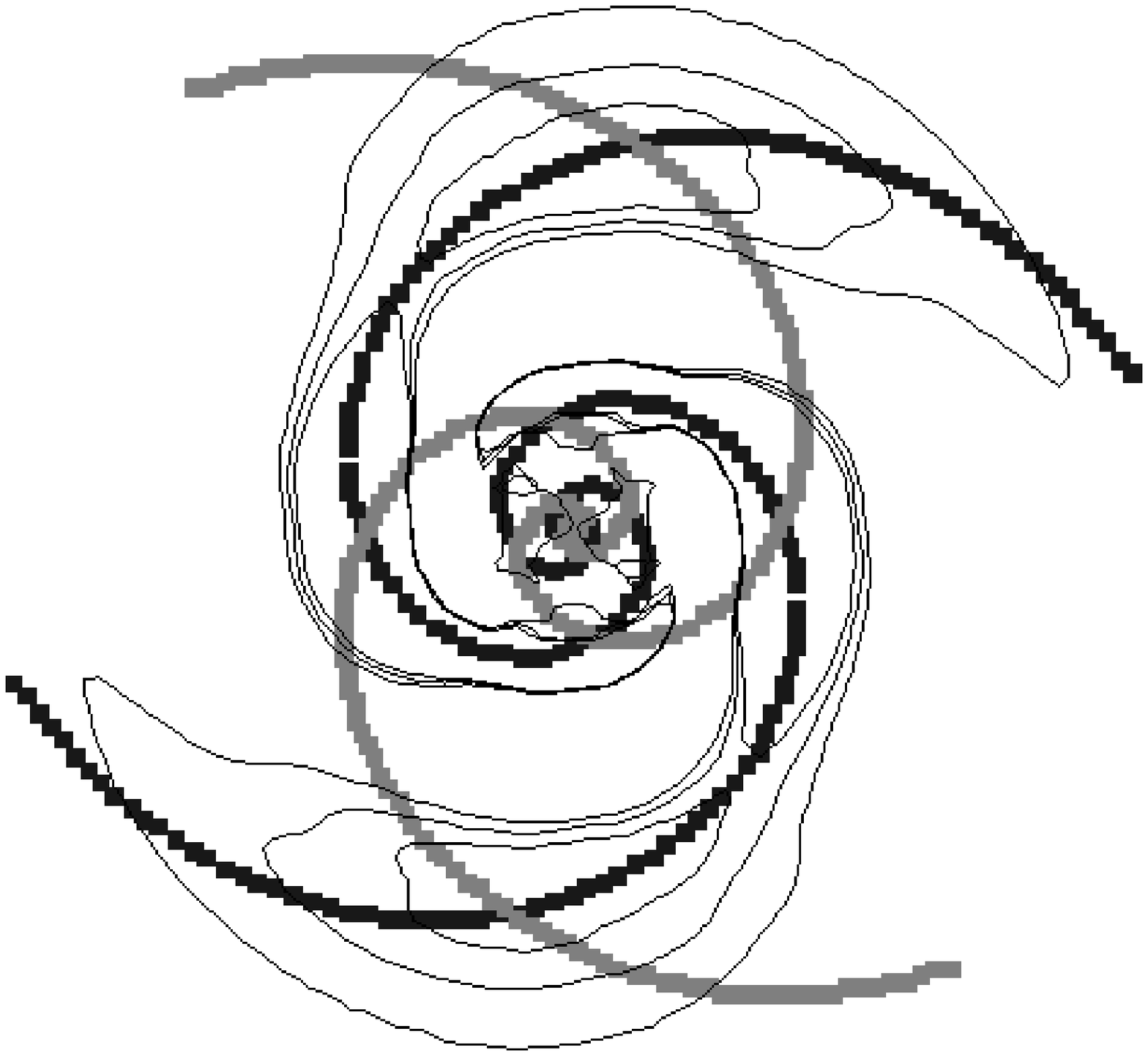}
          \caption{The synthetic logarithmic spirals, and the resulting
             contours for the $m$=2 mode. The galaxy is assumed to rotate
             in a clockwise direction. The dark spirals represent
             trailing arms, while leading arms are indicated by spirals
             in light grey. Note the striking similarity between the contours
             in this Figure and those in Figure \ref{imag4062}. The existence
             of the leading patterns can only indirectly be inferred from the
             modulation in the spiral arms.}
\label{imagmodel}
\end{figure}

\section{Discussion}

The existence of leading spiral patterns in a differentially
rotating disk is a complex issue.  Questions such as how such
patterns can survive the shearing of a disk differential rotation come to
the fore. How can one infer the presence of leading and trailing spiral
arms if they are not both directly seen?

We have constructed a simple model using the values of \object{NGC 4062}
from Table \ref{tabspec}. The model is very simple in the sense that we
do not give a radial dependence of arm density. The synthetic image has
a density equal to unity for the main trailing arms, and a density
equal to 0.65 for the weaker leading patterns (this was chosen to get
the same ratio $\vert A^L \vert/\vert A^T \vert$).

The Fourier spectra of these synthetic logarithmic spirals are shown in Figure
\ref{specmodel}. The synthetic spirals, together with the contours for
the $m$=2 component are shown in Figure \ref{imagmodel}. Note that the
leading patterns do not appear directly on the contours. {\it Nevertheless,
their existence can be inferred from the interference patterns}.
As expected, the contours show maxima at the intersection of the two
(trailing and leading) patterns.

Therefore, if the synthetic spirals represent incoming and outgoing
spiral density waves on an axisymmetric disk, an interference pattern
will be established where stellar density will be larger at locales of
constructive interference in our near-infrared images.

Although the phases of the two patterns are different (see Table
\ref{tabspec}), the maxima are separated by approximately
90$^{\circ}$. To quote Toomre (\cite{toomre81}), {\it ``The
90\,$^{\circ}$ spacing of their successive density maxima
[in his models] argues eloquently for the presence of trailing
and leading waves of very similar wavelengths, ...''}.

\section{Conclusions}

A two dimensional Fourier analysis provides a robust method for
detecting both trailing and leading spirals in galaxies far more
distant than \object{M51} or \object{M81}.

We have applied our method to near-infrared images of \object{NGC 4062}
(optically flocculent) and \object{NGC 5248} (optically grand design).
In both cases, we have inferred the existence of dominant
$m$=2 trailing and secondary $m$=2 leading spirals in the Fourier
spectra. The consequence of two peaks with {\it different
winding sense} in the $m$=2 component directly implies
spiral arm modulation.

In each case, the pitch angle of both trailing ($T$) and leading ($L$)
waves is almost the same ($P^T$=$19^\circ$ and $P^L$=$-24^\circ$ for
\object{NGC 4062}, and $P^T$=$28^\circ$ and $P^L$=$-30^\circ$ for
\object{NGC 5248}).  This was the case also for \object{M81}
(Elmegreen et al. \cite{elmelmsei89}). This is highly suggestive
that the incoming and the outgoing wave-trains have similar wavelengths.

The amplitude ratio $\vert A^L\vert / \vert A^T\vert$ is about 0.5
in both \object{NGC 4062} and \object{NGC 5248}, wherein the higher
amplitude is consistently assigned to the trailing mode. The amplitude
ratio is independent of absolute magnitude (--19.44 for \object{NGC 4062}
and --21.19 for \object{NGC 5248}) and arm class.

The arm/interarm contrast increases for both galaxies, but not
in a monotonic way. The sinusoidal behaviour (seen in the modulation
of the intensity) betrays the interference between incoming and
outgoing density waves.

This study has also demonstrated the
efficiency of near-infrared images for
understanding the mass distributions in galaxies
which appear quite dis-similar optically, but which
have much in common when examined in the infrared
regime.

Our observations of spiral arm amplitude modulations supports
the idea originally proposed by Lin (\cite{lin70}) that density waves turn
around by reflection or refraction inside galaxies, presumably in the
inner regions where the incoming, short-wavelengths, trailing waves
(Toomre \cite{toomre69}) meet the kinematically hot stellar bulge.
In the present study, the observed
symmetry of the leading waves, with pitch angles comparable to those
of the trailing waves, indicates that the outgoing waves also have
short-wavelengths, and this is consistent with the expected group
velocity of outward-moving, {\it leading} waves. This result implies
that the bulge region {\it reflects} incoming waves (no change of pitch angle),
but changes their sense of winding. Moreover, the amplification mechanism,
which is always at co-rotation, must be WASER type II, rather than WASER
type I. The WASER I mechanism involves outward-moving, {\it long}-wavelengths, {\it
trailing} waves formed by {\it refraction} near the bulge and amplified
at co-rotation without a change in winding sense, i.e. by superreflection
(Mark \cite{mark76, mark77}).

Modal analysis of the structures found here, using also the observable
velocity dispersions of the stars, could in principle determine the disk
and halo mass distributions and the pattern speeds of the spirals, as
was done for \object{M81} (Lowe et al. \cite{loweetal94}). Spiral arm amplitude
variations have the potential to become a powerful constraint for the
study of galactic dynamics.

\object{NGC 4062} represents the very first detection of spiral arm modulation
in the evolved stellar disk of an {\it optically flocculent} galaxy.

\begin{acknowledgements}

The authors are indebted to the Anglo-American Chairman's Fund Educational
Trust. A note of deep appreciation is expressed to Mrs M. Keeton and
the Board of Trustees. This research is partially supported by the Mexicain
Foundation CONACYT under the grant No. 28507-E. The OSU Bright Spiral Galaxy
Survey has been supported by NSF grants AST-9217716 and AST-9617006 to JAF.

\end{acknowledgements}

\end{document}